\documentclass[pra,superscriptaddress,twocolumn]{revtex4-2}

\usepackage{enumerate}
\usepackage{amsfonts,amssymb,amsmath}
\usepackage[]{graphics,graphicx,epsfig}
\usepackage{amsthm}
\usepackage{graphicx}
\usepackage{dcolumn}
\usepackage{natbib}
\usepackage{color}
\usepackage{multirow}
\usepackage{ulem}
\usepackage{bm}
\usepackage{url}

\usepackage[colorlinks = true, citecolor=blue, linkcolor=blue, urlcolor=blue]{hyperref}
\newcommand*{\fullref}[1]{\hyperref[{#1}]{\autoref*{#1} \nameref*{#1}}}

\def\ket#1{\left|#1\right\rangle}

\def\braket#1{\left\langle#1\right\rangle}

\newcommand{\tr}[1]{\text{Tr}\left[#1\right]}

\begin{document}


\title{Simulations of quantum non-locality with local negative bits}
\author{Kelvin Onggadinata}
\affiliation{Centre for Quantum Technologies,
National University of Singapore, 3 Science Drive 2, 117543 Singapore,
Singapore}
\affiliation{Department of Physics,
National University of Singapore, 3 Science Drive 2, 117543 Singapore,
Singapore}

\author{Pawel Kurzynski}
\affiliation{ Institute of Spintronics and Quantum Information, Faculty of Physics, Adam Mickiewicz University, Uniwersytetu Pozna{\'n}skiego 2, 61-614 Pozna\'n, Poland}
\affiliation{Centre for Quantum Technologies,
National University of Singapore, 3 Science Drive 2, 117543 Singapore,
Singapore}

\author{Dagomir Kaszlikowski}
\email{phykd@nus.edu.sg}
\affiliation{Centre for Quantum Technologies,
National University of Singapore, 3 Science Drive 2, 117543 Singapore,
Singapore}
\affiliation{Department of Physics,
National University of Singapore, 3 Science Drive 2, 117543 Singapore,
Singapore}

\date{September 6, 2023}


\begin{abstract}
We propose a simple simulation of nonlocal quantum correlations among $N$ qubits using a local hidden variable source with a positive probability distribution, given that each of the $N$ observers has access to a local negative bit. Notably, unlike the Toner-Bacon protocol, no exchange of classical bits between the observers is required. Moreover, our simulation can be extended to include Popescu-Rohrlich box correlations.
\end{abstract}

\maketitle

\section{Introduction}\label{sec: introduction}

Consider an experiment where a quantum source emits two qubits in some state $\rho_{AB}$ to two spatially separated observers, Alice and Bob. Each observer measures their qubit in two randomly chosen bases, given by unit Bloch vectors $\hat{a}_0$ and $\hat{a}_1$ for Alice and $\hat{b}_0$ and $\hat{b}_1$ for Bob. For each basis choice $\hat{a}_i$ and $\hat{b}_j$, Alice's and Bob's outcomes $a_i,b_j=\pm 1$ are distributed with probabilities $p(a_i,b_j|\hat{a}_i,\hat{b}_j)=\tr{\rho_{AB} P(a_i|\hat{a}_i)\otimes P(b_j|\hat{b}_j)}$, where $P(x|\hat{x})=\frac{1}{2}(1+x\hat{x}\cdot\hat{\sigma})$, $x=\pm 1$, is the standard projective measurement operator for a qubit in the $\hat{x}$ direction and $\hat{\sigma}=(\sigma_x,\sigma_y,\sigma_z)$ is the vector of Pauli operators. Throughout the paper we denote normalized vectors by circumflexes $\hat{a}$ and unnormalized ones by arrows $\vec{a}$. 

The local hidden variable (LHV) hypothesis, first proposed by Einstein, Podolsky, and Rosen \cite{einstein1935quantum}, postulates that the quantum source $\rho_{AB}$ can be replaced with a source emitting particles carrying deterministic information $\lambda$ of what outcome to produce for the randomly chosen bases. For instance, $\lambda=(a_0,a_1;b_0,b_1)$, instructs the particles to give outcomes $a_0$ for Alice's basis $\hat{a}_0$ and $a_1$ for her basis $\hat{a}_1$, and $b_0$ and $b_1$ for Bob's respective bases. We can easily see that only $16$ such instructions are needed, i.e., we have $\lambda_i$ ($i=0,1,2,\dots,15$). To account for quantum randomness, these deterministic instructions $\lambda_i$ must be distributed by the source with some positive joint probability distribution (JPD) $\rho(\lambda_i)=\rho(a_0,a_1;b_0,b_1)$ such that (i) $\rho(a_0,a_1;b_0,b_1)\geq 0$,  (ii) $\sum_{a_0,a_1,b_0,b_1} \rho(a_0,a_1;b_0,b_1)=1$, and (iii) the marginals $p(a_i,b_j)=\sum_{\not{a_i}, \not{b_j}}\rho(a_0,a_1;b_0,b_1)$ should reproduce quantum probabilities; here $\sum_{\not{a_i}, \not{b_j}}$ denotes a summation over the outcomes that are not $a_i$ and not $b_j$. Note that the instructions $\lambda_i$ can be viewed as bit strings if we identify $-1\rightarrow 0$ and $+1\rightarrow 1$. This observation will be used later in the paper.

Bell proved \cite{bell1964einstein} that there are entangled states $\rho_{AB}$ and choices of measurement bases such that observed quantum probabilities $p(a_i,b_j|\hat{a}_i,\hat{b}_j)$ cannot be simulated with LHVs distributed via some JPD $\rho(a_0,a_1;b_0,b_1)$ if (i)--(iii) are satisfied. Although he showed it for a two-qubit singlet state, other researchers followed with sweeping generalizations for an arbitrary number of qubits, measurement settings, and higher-dimensional quantum states \cite{mermin1990extreme,zukowski2002bell,sliwa2003symmetries}. Subsequently, it was noticed that if one relaxes (i) and admits a joint quasiprobability distribution (JQD), LHV simulations are possible \cite{abramsky2004operational,alsafi2013simulating}. Let us give a simple example. 

Consider a singlet state $|\psi_-\rangle_{AB}$ and measurement settings $\hat{a}_0=\hat{x}$ and $\hat{a}_1=\hat{z}$ for Alice and $\hat{b}_0=\frac{1}{\sqrt{2}}(\hat{x}+\hat{z})$ and $\hat{b}_1=\frac{1}{\sqrt{2}}(\hat{x}-\hat{z})$ for Bob. They yield a simple set of quantum probabilities $p(a_i,b_j|\hat{a}_i,\hat{b}_j)=\frac{1}{4}(1-m_{ij}\frac{a_ib_j}{\sqrt{2}})$, where $m_{ij}=1-2\delta_{i,1}\delta_{j,1}$. These probabilities cannot be simulated with any JPD. However, the following JQD mimics these probabilities perfectly:
\begin{equation}
\rho(a_0,a_1;b_0,b_1)=\frac{1}{16}\left(1-\sum_{i,j}\frac{m_{ij}}{\sqrt{2}}a_ib_j\right)\,.
\end{equation}
Note that some joint probabilities $\rho(a_0,a_1;b_0,b_1)$ are negative but only the positive marginals $p(a_i,b_j)$ can be observed. This constraint defines the rules of a general simulation game: Negative probabilities can never appear for probabilistic events we can observe in the laboratory.

Abramsky and Brandenburger \cite{abramsky2004operational} showed that quantum probabilities $p(a_i,b_j|\hat{a}_i,\hat{b}_j)$ can always be simulated with a JQD if a JPD simulation is not possible. We need to stress here that they place negativity necessary for the simulation right in the LHV source, replacing the quantum state $\rho_{AB}$. In this paradigm, quantum measurements on Alice's and Bob's sides are direct readouts of instructions carried by the LHVs, i.e., if Alice chooses to measure in the basis $\hat{a}_i$ and Bob in $\hat{b}_j$, they get $a_i$ and $b_j$ from the distributed LHV variable $\lambda=(a_0,a_1;b_0,b_1)$. Because of the complementarity and the irreversible nature of the measurement process, they ignore the rest of the information contained in $\lambda$.  

Later Al-Safi and Short \cite{alsafi2013simulating} reproduced the Abramsky-Brandenburger result and also provided a proof of the principle that all no-signaling correlations can be simulated with an LHV source distributed with a JPD and local measurement strategies with negative probabilities. They were not concerned about the cost of their simulation.  

Toner and Bacon \cite{toner2003communication} considered a different simulation protocol for bipartite quantum correlations for qubits. They use a source of LHVs distributed with a JPD but Alice and Bob have to exchange approximately $0.85$ bits of classical communication on average to simulate quantum correlations. They trade negativity in a JQD for an exchange of classical bits, setting a different paradigm from the previous one.  
 
Here we propose a concrete algorithm implementing Al-Safi and Short's paradigm for an LHV simulation that recovers all observable quantum probabilities between $N$ qubits in an arbitrary quantum state $\rho_{A_1A_2\ldots A_N}$ generated by $N$ spatially separated observers $A_1, A_2,\dots , A_N$, each measuring their qubits with an arbitrary number of measurement settings. Similarly to Toner and Bacon's simulation, ours uses a source of LHVs distributed with a JPD (not a JQD) but we replace the exchange of classical bits with local negative bits used to locally process the observers' LHV data. Unlike in Al-Safi and Short's approach, we can optimize the amount of local negativity and thus find a cost of nonlocality simulation.

\section{Negative bit}\label{sec: nebit}

We introduce here a negative bit, previously discussed in \cite{kaszlikowski2021little}. It is a binary system whose values $n=\pm 1$ appear with quasiprobabilities $w(n)=\frac{1}{2}(1+\frac{n}{\lambda})$ where $|\lambda|<1$ (if $|\lambda|\geq 1$ we have a non-negative random bit). If $0<\lambda<1$ then $w(+1)>1$ and we call it an inflated probability, while $w(-1)<0$ is a negative probability. Similarly if $-1<\lambda<0$ then $w(-1)$ is inflated and $w(+1)$ is negative. 

A negative bit is a natural unit of quasiprobability since every quasiprobability distribution $\{p_1,p_2,\ldots,p_k,q_{1},\ldots,q_{n}\}$, where $p_i \geq 0$ and $q_j <0$ can be written as
\begin{eqnarray}
& &\frac{w(+1)}{\sum_{i=1}^k p_i}\{p_1,p_2,\ldots,p_k,0,\ldots,0\} \nonumber \\
&+&\frac{w(-1)}{\sum_{j=1}^n |q_j|}\{0,0,\ldots,0,|q_{1}|,\ldots,|q_{n}|\}
\end{eqnarray}
for $w(+1)=\sum_{i=1}^k p_i$ and $w(-1)=\sum_{j=1}^n q_j$. From this we can evaluate
\begin{equation}
\lambda = \frac{1}{2\left(\sum_{i=1}^k p_i\right)-1}.
\end{equation}
This negative bit decomposition is significant when one deals with quasi-bistochastic processes that are quasiprobabilisitic versions of bistochastic processes as discussed in detail in Appendix \ref{sec: appendix bvn decomposition}.

\section{Simulation}\label{sec: simulation}

\subsection{Two parties}

\begin{figure}[!htb]
    \centering
    \includegraphics[width=1.0\linewidth]{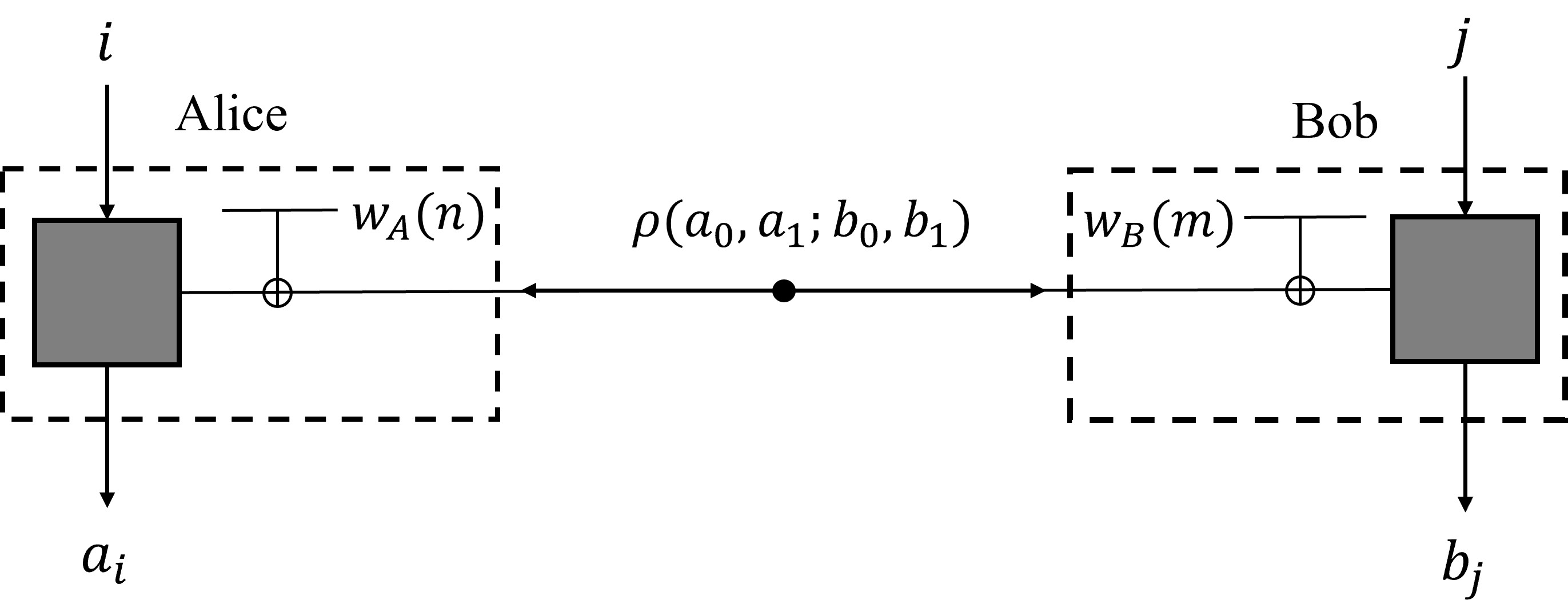}
    \caption{Simulation of quantum measurements with negative bits. A local hidden variable source distributes instructions $(a_0,a_1;b_0,b_1)$ to local observers Alice and Bob. Each instruction to generate bits $a_i$ and $b_j$ if Alice chooses to measure in the basis $x=i$ and Bob in the basis $y=j$ is distributed with a joint positive probability distribution $\rho(a_0,a_1;b_0,b_1)$ that depends on to-be-simulated quantum state $\rho$. Alice and Bob's measuring apparatus execute a CNOT gate on the incoming bits $a_i$ and $b_j$, controlled by local negative bits $n$ and $m$, each generated with a negative probability $w_A(n)$ and $w_B(m)$. The statistics of the outcomes $a_j$ and $b_j$ faithfully reproduces quantum mechanical measurement probabilities for the bases $i$ and $j$ and the state $\rho$.}
    \label{fig: nebit simulation}
\end{figure}

Let us start with the singlet and two measurement settings for the Alice and Bob example we described in the Introduction. We assume for now that the source produces LHVs with a JPD:
\begin{eqnarray}
&&\rho(a_0,a_1;b_0,b_1)\nonumber\\ 
&& = \frac{1}{16}\left[1-\frac{1}{2}(a_0b_0+a_0b_1+a_1b_0-a_1b_1)\right]\, .
\label{sim0}
\end{eqnarray}
Unlike the JQD we used before, the above distribution is always positive and thus it cannot reproduce the quantum probabilities, giving us $\rho(a_i,b_j)=\frac{1}{4}(1-\frac{1}{2}m_{ij}a_ib_j)\geq 0$. We are missing the right factor in front of $m_{ij}a_ib_j$, which should be $\frac{1}{\sqrt{2}}$ for a faithful mimicry. This is not a problem if we realize that Alice and Bob use a measuring apparatus in the laboratory. Such an apparatus is a device that amplifies and irreversibly records a signal triggered by a microscopic entity we call a qubit. Here, the qubit is represented by the LHVs and so for a successful simulation we need to design a proper measuring apparatus. The simplest choice is a controlled-NOT (CNOT) gate controlled by a negative bit as introduced in \cite{kaszlikowski2021little} with the negative binary distribution (the same for Alice and Bob because of the system's symmetries) $w(n)=w_A(n)=w_B(n)=\frac{1}{2}(1+n\sqrt{\sqrt{2}})$, where $n=\pm 1$ (see Fig. \ref{fig: nebit simulation}). For instance, Alice and Bob's measurement probabilities in the bases $\hat{a}_0$ and $\hat{b}_1$ are faithfully recovered:
\begin{eqnarray}
&&p(a_0,b_1)\nonumber\\
&& = \sum_{a_1,b_0,n_0, m_1}\rho(a_0n,a_1;b_0,b_1m)w_A(n)w_{B}(m).
\end{eqnarray}
The above formula is clear if one notices that in our notation $G_{\text{CNOT}}[p(a)w(n)]:=p(an)w(n)$, i.e., $a=\pm 1$ is multiplied by $n=\pm 1$. This is equivalent to XOR for bits represented by $0 \rightarrow 1$ and $1 \rightarrow -1$. Since we do not register all variables, we sum over those we do not measure. In addition, another property of our quantum measurement simulation is complementarity, i.e., Alice and Bob must commit to a measurement basis because no one knows how to build an apparatus that could measure two or more complementary observables simultaneously. Complementarity guarantees that Alice and Bob negative bits' negativities are never directly observed. 

This simulation easily extends to an arbitrary state $\rho_{AB}$, given by local Bloch vectors $\vec{s}_A$ and $\vec{s}_B$, the correlation matrix $T_{AB}$, and two arbitrary measurement bases for Alice and Bob. We build a  positive LHV distribution $\rho(a_0,a_1;b_0,b_1)$ as 
\begin{eqnarray}
&&\rho(a_0,a_1;b_0,b_1) \nonumber\\
&& = \frac{1}{16}\Bigg(1+\lambda\sum_{i=0}^1 a_i\langle A_i\rangle+\lambda\sum_{j=0}^1 b_j\langle B_j\rangle\nonumber\\
&& \quad + \lambda^2\sum_{i,j=0}^1a_ib_j\langle A_iB_j\rangle \Bigg)\, , \label{distr}
\end{eqnarray}
where $\langle A_i\rangle=\hat{a}_i\cdot\vec{s}_A$ and $\langle B_j\rangle=\hat{b}_j\cdot\vec{s}_B$ are the first-order and $\langle A_iB_j\rangle = \hat{a}_i\cdot T_{AB}\cdot \hat{b}_j$ the second-order quantum-mechanical correlation functions. In the above $\vec{s}_X$ $(X=A,B)$ is a local Bloch vector of the corresponding qubit and $T_{AB}$ is a correlation tensor of the two-qubit system. 

The positivity of this distribution is guaranteed by a proper choice of $\lambda$. In this case, the $\lambda$ can be found by making sure that all CHSH inequalities are satisfied, i.e., 
\begin{equation}
\lambda^2(|\langle A_0B_0\rangle +\langle A_0B_1\rangle|+|\langle A_1B_0\rangle -\langle A_1B_1\rangle|)\leq 2\, ,
\end{equation}
giving us
\begin{equation}
|\lambda| \leq \sqrt{ \frac{2}{|\langle A_0B_0\rangle +\langle A_0B_1\rangle|+|\langle A_1B_0\rangle -\langle A_1B_1\rangle|}}\, .
\end{equation}
If we do not care to get the largest possible $\lambda$, we can simply grossly underestimate the lower bound of Eq. (\ref{distr}) as
\begin{eqnarray}
&&\rho(a_0,a_1;b_0,b_1) \nonumber\\
&&\geq \frac{1}{16}\Big[1-2\lambda (s_A+s_B) - 4\lambda^2 \Vert T_{AB}\Vert \Big]\, , 
\end{eqnarray}
where $s_A=|\vec{s}_A|$, $s_B=|\vec{s}_B|$, and $\Vert T_{AB}\Vert =\sqrt{\sum_{m,n}(T_{AB})_{nm}^2}$. Now we demand that this lower bound is non-negative. In particular, if it is zero, we get a quadratic equation for $\lambda$, from which we find
\begin{equation}
\lambda = \frac{-2(s_A+s_B)+\sqrt{(2s_A+2s_B)^2+16\Vert T_{AB}\Vert^2}}{8\Vert T_{AB}\Vert }\, .    
\end{equation}

The measurement apparatuses are, like before, the CNOT gates with a negative bit $w(n)=\frac{1}{2}(1+\frac{n}{\lambda})$ for both Alice and Bob. We can see that the largest possible $\lambda$ minimizes the negative bit's negativity $\frac{1}{2}(1-\frac{1}{\lambda})$ and it matters if we are interested in the cost of the simulation. If we are not, the suboptimal $\lambda$ shown above will do.

We remark that negative bits appear in quantum theory so they are not something entirely exotic. To see it, let us consider a symmetric, informationally complete, positive operator-valued measure quasiprobability representation of a qubit \cite{filippov2010symmetric,kiktenko2020probability}: (i) The qubit's density matrix $\rho$ with a Bloch vector $\vec{s}$ is represented as a positive probability distribution $\rho_k=\frac{1}{4}(1+\hat{n}_k\cdot\vec{s})$, $k=1,2,3,4$, where the $\hat{n}_k$ are tetrahedron spanning vectors, i.e., $\sum_k\hat{n}_k=0$, $\hat{n}_k\cdot\hat{n_l}=\delta_{kl}-\frac{1}{3}(1-\delta_{kl})$. (ii) Unitary operations are represented by quasi-bistochastic matrices $S = [S_{kl}]_{kl}$ with its elements given by $S_{kl} = \frac{1}{4}+\frac{3}{4}O(U)\hat{n}_k\cdot\hat{n_l}$, where $O(U)$ is the three-dimensional orthogonal representation of a unitary $U$. Now, any such $S$ can be represented as two positive bistochastic processes $S^{0}$ and $S^{1}$ controlled by a suitably chosen negative bit, i.e., the process $S^{0}$ is activated with the probability $\frac{1}{2}(1+\eta)$ and $S^{1}$ with the probability $\frac{1}{2}(1-\eta)$, where $\eta$ is greater than one if $S$ is not a permutation matrix (see Appendix \ref{sec: appendix bvn decomposition} for details). (iii) Quantum measurement is represented as an effect $\vec{m}_a$ that is a suitably chosen linear combination of the $\hat{n}_k$. The measurement probability is then obtained via $p(a|\vec{a})=\vec{m}_a\cdot \vec{\rho}$, where $\vec{\rho}=[\rho_1,\rho_2,\rho_3,\rho_4]$. Thus, this quasiprobability representation of a qubit's mechanics can be viewed as an example of a negative bit simulation. The crucial difference is that in our simulation we do not need effects as the measurement outcomes are directly encoded in the initial probability distribution.   

Finally, we need to stress out that the negative bits used in the simulation are local, no exchange of classical bits is necessary, and the whole model is a no-signaling one. This remark is related to the Toner-Bacon model \cite{toner2003communication}, where Alice and Bob simulate bipartite quantum correlations with LHVs and an exchange of one bit of classical information.  

Our simulation can be easily extended to Popescu-Rohrlich (PR) box correlations \cite{popescu1994quantum}. We start with the same positive distribution we used to simulate the singlet correlations (\ref{sim0}) but increase the negative bit's negativity to $w(n) = \frac{1}{2}(1 + n\sqrt{2})$. This procedure can be pushed farther to extend quantum mechanical correlations for an arbitrary two-qubit state $\rho_{AB}$ beyond quantum theory (see Appendix \ref{sec: appendix nebit pr-box}).

\subsection{$N$ parties}

Extension to an arbitrary number of Alice's and Bob's local measurements $\hat{a}_i$ ($i=0,1,\dots, N_A-1$) and $\hat{b}_j$ ($j=0,1,\dots, N_B-1$) is straightforward. The initial positive LHV distribution $\rho(a_0,a_1,\dots, a_{N_A-1};b_0,b_1,\dots, b_{N_B-1})$ is 
\begin{eqnarray}
&&\rho(a_0,a_1,\dots, a_{N_A-1};b_0,b_1,\dots, b_{N_B-1})\nonumber\\
&& = \frac{1}{2^{N_A+N_B}}\Bigg(1+\lambda\sum_{i=0}^{N_A-1}a_i\langle A_i\rangle+\lambda\sum_{j=0}^{N_B-1}b_j\langle B_j\rangle \nonumber\\
&& \qquad +\lambda^2\sum_{i=0}^{N_A-1}\sum_{j=0}^{N_B-1}a_ib_j\langle A_iB_j\rangle\Bigg).
\end{eqnarray}
The largest possible $\lambda$ ($|\lambda|\leq 1$) for which this distribution is positive can be obtained numerically. In any case, setting
\begin{eqnarray}
\lambda &=& \frac{1}{2N_AN_B\Vert T_{AB}\Vert}\Big[-s_AN_A-s_BN_B \nonumber\\
& & \quad +\sqrt{(s_AN_A+s_BN_B)^2+4N_AN_B\Vert T_{AB}\Vert }\Big]
\end{eqnarray}
suffices to make it a positive LHV distribution although this $\lambda$ is grossly suboptimal (small).
Measuring apparatuses are the same as before, i.e., local CNOT gates with the negative bit $w(n)=\frac{1}{2}(1+\frac{n}{\lambda})$. Another way to understand this proof is to observe that the CNOT gate with the negative bit $w(n)$ can reverse noise $\lambda$ on any probability distribution $p(a)=\frac{1}{2}(1+a\lambda\langle A\rangle)$, i.e., $\sum_{n}p(an)w(n)=\frac{1}{2}(1+a\langle A\rangle)$. 

We now consider $N$ qubits measured by $N$ observers, where the $k$th ($k=0,1,\dots, N-1$) observer has $N_k$ different measurement directions $\hat{a}^{(k)}_{i_k}$ ($i_k=0,1, \dots, N_k-1$). The simulation protocol starts with a positive $N$-party LHV distribution
\begin{eqnarray}
&&\rho(a^0_0, a^0_1,\dots, a^0_{N_0-1};\dots; a^{N-1}_0, a^{N-1}_1,\dots, a^{N-1}_{N_{N-1}-1})\nonumber\\
&&= \frac{1}{2^{N_0+N_1+\dots N_{N-1}}}\Big(1+\lambda\sum_{k=0}^{N-1}\sum_{i_k=0}^{N_k-1}a^{(k)}_{i_k}\langle A^{(k)}_{i_k}\rangle\nonumber\\
&& + \lambda^2\sum_{k\neq l=0}^{N-1}\sum_{i_k=0}^{N_k-1} \sum_{i_l=0}^{N_l-1}  a^{(k)}_{i_k}a^{(l)}_{i_l}\langle A^{(k)}_{i_k}A^{(l)}_{i_l}\rangle+\dots \nonumber\\ &&+\lambda^N\sum_{i_0,i_1,\dots, i_{N-1}=0}^{N_0,N_1,N_{N-1}}a^0_{i_0}\dots a^{N-1}_{i_{N-1}}\langle A^{(0)}_{i_0}\dots A^{(N-1)}_{i_{N-1}}\rangle \Big)\, ,
\end{eqnarray}
where $\lambda$ is chosen to make it positive. This requires finding roots of an $N$th degree polynomial and it can only generally be done numerically. Since for $\lambda=0$ the distribution is positive and for $\lambda =1$ it can be negative, there must exist a range of $\lambda$ for which the distribution is positive as the problem is continuous in $\lambda$. As long as this $\lambda$ is strictly positive, a CNOT gate with the negative bit $w(n)=\frac{1}{2}(1+\frac{n}{\lambda})$ for each observer will recover quantum-mechanical measurement probabilities.

We illustrate this with a Mermin inequality for ($N=3$ and $N_k=2$, $k=0,1,2$) \cite{mermin1990extreme}. One possible form of the Mermin inequality is
\begin{equation}
    M = a_0b_0c_0 - a_0b_1c_1 - a_1b_0c_1 - a_1b_1c_0
\end{equation}
and the maximal value achieved by any JPD $\rho(a_0,a_1;b_0,b_1;c_0,c_1)$ is $\vert\langle M \rangle_{LHV}\vert \leq 2$. An instance of such a JPD is
\begin{eqnarray}
&& \rho(a_0,a_1;b_0,b_1;c_0,c_1)  \nonumber \\
&& = \frac{1}{2^6}\Big[1 + \frac{1}{2}(a_0b_0c_0  + a_0b_1c_1 - a_1b_0c_1 - a_1b_1c_0)\Big]\, .
\end{eqnarray}
Implementing the local CNOT gates with each party using negative bit $w(n) = \frac{1}{2}(1+ n2^{1/3})$ yields the measurement probabilities
\begin{eqnarray}
&& p(a_i,b_j,c_k)  \nonumber \\
&& = \sum_{\substack{\not{a_i},\not{b_j},\not{c_k}\\n_A,n_B,n_C}}G_{\text{CNOT}}[\rho(a_0,\dots,c_1),w_A(n_A)w_B(n_B)w_C(n_C)] \nonumber \\
&&=\sum_{n_A,n_B,n_C}G_{\text{CNOT}}[\rho(a_i,b_j,c_k),w_A(n_A)w_B(n_B)w_C(n_C)] \nonumber \\
&& =\sum_{n_A,n_B,n_C}\rho(n_Aa_i,n_Bb_j,n_Cc_k)w_A(n_A)w_B(n_B)w_C(n_C)\, \nonumber \\
\end{eqnarray}
It can be check easily that this corresponds to the quantum measurement probabilities on the state $|\mathcal{S}_{\text{GHZ}_3}\rangle = \frac{1}{\sqrt{2}}(\ket{000} + \ket{111)})$ and measurement settings $\hat{a}_0=[1,0,0]$, $\hat{b}_0=\hat{c}_0=[-1,0,0]$, $\hat{a_1}=[0,1,0]$,and $\hat{b}_1=\hat{c}_1=[0,-1,0]$ achieving a violation of $\vert \braket{M}_Q\vert=4$. Generalization to any forms of Bell-type inequalities follow the same idea.

\section{Discussion}

In this paper we have focused on $N$-qubit correlations generated by $N$ spatially separated observers, each measuring an arbitrary number of complementary observables. We have explicitly demonstrated how to simulate this setup using (i) a source dispatching local hidden variables with positive probabilities and (ii) a logical CNOT gate controlled by a local negative bit. 

This local negative bit modifies the JPD of LHVs to replicate faithfully quantum-mechanical measurements or even PR boxes, given the availability of a sufficient amount of local negativity. We need to stress that, unlike in the Toner-Bacon model \cite{toner2003communication}, we do not require an exchange of classical bits between observers. It is an open question how the negative bit's mathematical negativity relates to the amount of physical classical bits in the Toner-Bacon model. In order to make a meaningful comparison it is necessary to minimize the amount of negativity of each local negative bit. This is not a trivial task but it can be accomplished numerically if required. However, a more in depth enquiry is necessary to find this connection, which extends beyond the scope of this paper. 

We would like to stress that negative bits appear naturally in quasiprobability representations of quantum mechanics as we pointed out in Sec. \ref{sec: simulation}. However, our simulation uses them in a different way. The situation is similar to simulations in \cite{abramsky2004operational, alsafi2013simulating}, where the JQDs used are not equivalent to discrete Wigner-Wootters functions \cite{ferrie2009framed,ferrie2011quasi}. This different usage of the negative bit allows us to simulate PR boxes that extend beyond quantum theory.    

Note that in our simulation we never ``see" negative probabilities just like we never see them in quantum theory. They are hidden in the measuring apparatus and this is an important feature of our simulation because so far no one has found a commonly accepted operational meaning of negative probabilities (see, for instance, \cite{abramsky2004operational}).   

\section*{Acknowledgements}

This research was supported by the National Research Foundation, Singapore, and A*Star under the CQT Bridging Grant. P.K. was supported by the Polish National Science Centre (NCN) under the Maestro Grant No. DEC-2019/34/A/ST2/00081.

\bibliography{references}

\begin{appendix}

\section{Generalized Birkhoff--von Neumann decomposition of a quasi-bistochastic matrix}\label{sec: appendix bvn decomposition}

Here we show, using Birkhoff-von Neumann (BvN) decomposition \cite{valls2021birkhoff}, how to decompose any $d\times d$ quasi-bistochastic matrix $S = [S_{kl}]_{k,l=1}^d$ to two non-negative bistochastic processes controlled by a negative bit. A quasi-bistochastic matrix is a quasiprobabilistic generalization of a bistochastic matrix. Entries of a bistochastic matrix are non-negative and all rows and columns sum to one. Analogously, entries of a quasi-bistochastic matrix can be any real numbers, but all rows and columns must still sum up to one: $S_{kl}\in \mathbb{R}, \, \sum_k S_{kl}=1\, \forall l$, and $\sum_{l}S_{kl}=1\, \forall k$.

First we find the bistochastic matrix from $S$,
\begin{equation}
    B = \frac{1}{1 + d\Delta} \left(S + \Delta \mathbf{1}\right)\, ,
\end{equation}
where $\mathbf{1}$ is the matrix made of all ones and $\Delta = \max
\{0,-\min_{kl}\{S_{kl}\}\}$. This gives us
\begin{equation}
    S = (1+d\Delta)B - \Delta \mathbf{1}\,.
\end{equation}
Next we apply the BvN algorithm to $S$ and $\mathbf{1}$ to obtain
\begin{equation}\label{eq: decomposition of B and 1}
    B = \sum_{i=1}^N p^B(i) \Pi^B_i\,,  \quad \mathbf{1} = \sum_{j=1}^r \Pi^\mathbf{1}_j\,,
\end{equation}
where $\Pi_i$ is some permutation matrix and $p=[p(i)]_i$ is a positive probability distribution satisfying $\sum_i p(i) =1$. Here the superscript in $\{\Pi^B_i\}$ and $\{\Pi^{\mathbf{1}}_j\}$ is to clarify which decomposition it originates from. Note that the number of decomposition terms $N$ is smaller than $d^2$. 

From (\ref{eq: decomposition of B and 1}) we get our generalized BvN decomposition
\begin{equation}
\begin{split}
    S & = \sum_{i=1} q^S(i) \Pi^S_i \\ & = \sum_{j} q^{S,+}(j) \Pi^{S,+}_j - \sum_{k} |q^{S,-}(k)| \Pi^{S,-}_k\, ,
\end{split}
\end{equation}
where $q^S(i) \in \mathbb{R}$ and $\sum_i q^S(i) = \sum_j q^{S,+}(j) + \sum_k q^{S,-}(k) = 1$. We group $q^S(i)$ into $\{q^{S,+}(j)\}$ if it is positive and $\{q^{S,-}(k)\}$ if it is negative. This gives us
\begin{equation}
S = n^+ S^+ + n^- S^-\, ,
\end{equation}
where
\begin{equation}
n^\pm = \sum_k q^{S, \pm}(k)\, , \quad\,n^+ + n^- =1\, ,
\end{equation}
and
\begin{equation}
    S^{\pm} = \frac{1}{n^\pm}\sum_k q^{S,\pm}(k) \Pi^{S,\pm}_k\, .
\end{equation}
Note that $S^{\pm}$ are positive bistochastic matrices. Since the negative bit is the source of $S$'s negativity, we measure it as
\begin{equation}
    \mathcal{N} = |n^-|\, 
\end{equation}
and call it the negative bit's negativity. The maximum negativity happens for a decomposition with nonoverlapping $\{\Pi^B_j\}$ and $\{\Pi^\mathbf{1}_k\}$. In that case, the maximal negativity is $d\Delta$. If $S$ represents a unitary $U$, the upper bound for $\Delta$ can be calculated analytically \cite{zhu2016quasiprobability}. This gives the upper bound on $\mathcal{N}$ as well.

\section{Nebit simulation on nonmaximally entangled state}\label{sec: appendix nebit pr-box}

Here, we show that one can use the simulation protocol to extend any quantum mechanical correlations of an arbitrary two-qubit state $\rho_{AB}$ to the maximal post-quantum correlations. Consider a pure quantum state $\ket{\psi}=\alpha\ket{01} - \beta\ket{10}$, where $\alpha,\beta\geq 0$ and $\alpha^2+\beta^2 =1$. The method to find the optimal settings for the maximal violation of Bell-CHSH inequality can be found in \cite{horodecki1995violating}. For completeness, we show explicitly the direction of the optimal settings
\begin{equation}
    \begin{split}
        \hat{a}_0 &= [0, 0, -1]\, , \\
        \hat{a}_1 &= [-1, 0, 0]\, , \\
        \hat{b}_0 &= [\sin\theta, 0, \cos\theta]\, , \\
        \hat{b}_1 &= [-\sin\theta, 0, \cos\theta]\, ,
    \end{split}
\end{equation}
where $\theta = \arctan|2\alpha\beta|$, and correspondingly the local averages and two-point correlations
\begin{equation}
\begin{split}
\braket{A_0} &= \beta^2 - \alpha^2\, , \\
\braket{A_1} &= 0\, ,\\
\braket{B_0} &= \frac{\beta^2-\alpha^2}{\sqrt{1+4\alpha^2\beta^2}}\, , \\
\braket{B_1} &= \frac{\beta^2-\alpha^2}{\sqrt{1+4\alpha^2\beta^2}}
\end{split}
\end{equation}
and
\begin{equation}    
\begin{split}
    \braket{A_0B_0} &= \frac{1}{\sqrt{1+4\alpha^2\beta^2}} \, ,\\ 
    \braket{A_0B_1} &= \frac{1}{\sqrt{1+4\alpha^2\beta^2}} \, ,\\
    \braket{A_1B_0} &= \frac{4\alpha^2\beta^2}{\sqrt{1+4\alpha^2\beta^2}} \, ,\\
    \braket{A_1B_1} &= - \frac{4\alpha^2\beta^2}{\sqrt{1+4\alpha^2\beta^2}} \,.
    \end{split}
\end{equation}
Note that $A_k = \hat{a}_k\cdot \vec{\sigma}$, $\vec{\sigma} = [\sigma_x, \sigma_y, \sigma_z]$, is the spin operator with the $k$th setting. Consequently, the maximal quantum violation yields $Q:=2\sqrt{1+4\alpha^2\beta^2}\geq 2$. The quantum pair probabilities that saturate $Q$ read
\begin{equation}
    p(a_i, b_j)=\frac{1}{2^2}\left[1 + a_i\braket{A_i} + b_j\braket{B_j} + a_ib_j\braket{A_iB_j}\right]\, .
\end{equation}

\begin{figure}[!t]
    \centering
    \includegraphics[width=1.0\linewidth]{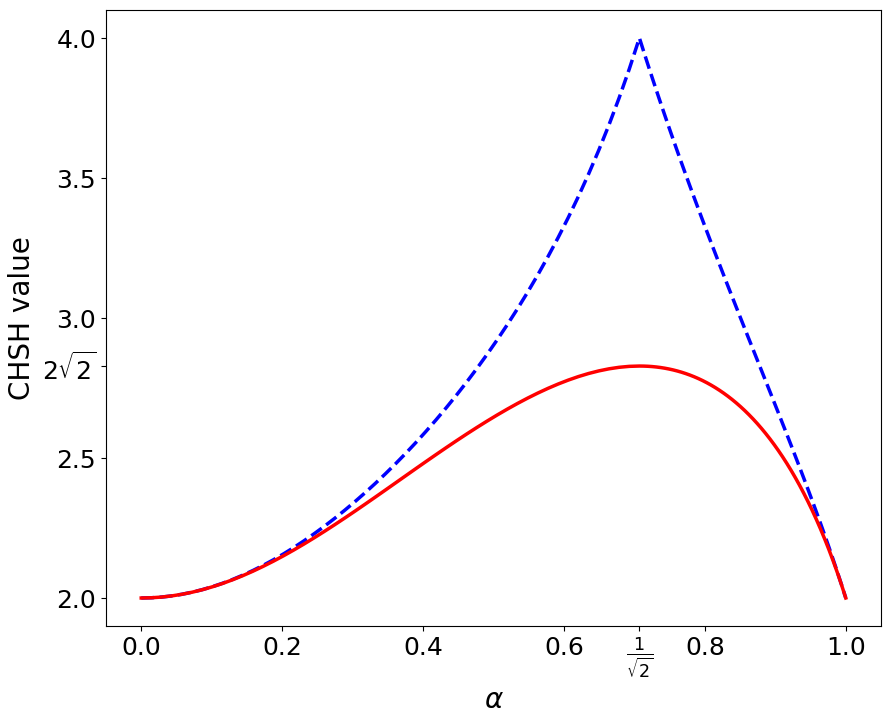}
    \caption{Maximal Bell-CHSH value attainable by non-maximally entangled pure state. The solid red and dashed blue line indicates the quantum bound and no-signaling (PR-box) bound, respectively}
    \label{fig: no signalling boundary}
\end{figure}

As mentioned in the main text, the upgrade is conducted locally by performing a CNOT gate with the negative bit $w_A(n) = w_B(n) = w(n) = \frac{1}{2}(1+ n\eta)$. The resulting pair probabilities then yield
\begin{equation}\label{eq: PR pair probabilities}
p_{\eta}(a_i, b_j)=\frac{1}{2^2}\left[1 + \eta a_i\braket{A_i} + \eta b_j\braket{B_j} + \eta^2 a_ib_j\braket{A_iB_j}\right]\, ,
\end{equation}
which gives us the PR boxlike distribution when $\eta>1$. Thus, we have the pair-probabilities that could take the correlation beyond the quantum bound by a factor of $\eta^2$ ($\eta>1$), i.e., $\eta^2 Q$. It is easy to find the largest $\eta$ such that $\eta^2 Q$ is maximized while satisfying the positivity condition of the pair probabilities (\ref{eq: PR pair probabilities}), as this will give us the boundary of the PR box in the no-signalling polytope \cite{brunner2014nonlocality}. We plot this maximal violation in Fig. \ref{fig: no signalling boundary}. As seen, the peak is achieved at $\alpha=\frac{1}{\sqrt{2}}$ (maximally entangled state) with quantum behavior reaching a Tsirelson bound of $2\sqrt{2}$ and PR box reaching 4.

\end{appendix}

\end{document}